\documentclass[aps,prb,twocolumn,showpacs,superscriptaddress]{revtex4}
\usepackage{amssymb}
\usepackage{graphicx}

\begin{document}

\title{Magnetoconductivity oscillations induced by intersubband excitation in a degenerate 2D electron gas}
\author{Yu.P. Monarkha}
\email[E-mail: ]{monarkha@ilt.kharkov.ua}
\affiliation{B. Verkin Institute for Low Temperature Physics and Engineering, 47 Nauky Ave.,
61103, Kharkiv, Ukraine}

\begin{abstract}
Magnetoconductivity oscillations and absolute negative conductivity
induced by nonequilibrium populations of excited subbands in a degenerate multisubband
two-dimensional electron system are studied theoretically. The displacement from
equilibrium, which can be caused by resonant microwave excitation or by any other reason,
is assumed to be such that electron distributions can no longer be described by a single Fermi
level. In this case, in addition to the well-known conductivity peaks
occurring at the Shubnikov-de Haas conditions and small peaks of normal intersubband
scattering, sign-changing oscillations with a different shape are shown to be possible. We
found also that even a small fraction of electrons transferred to the
excited subband can lead to negative conductivity effects.
\end{abstract}




\maketitle

\section{Introduction}

The transport properties of a 2D electron gas in a perpendicular
magnetic field have attracted much interest~\cite{AndFowSte-1982,PraGir-1987}
because of unexpected discoveries and new physics. In addition to the
amazing quantum Hall effects observed in a degenerate 2D electron gas under
equilibrium conditions~\cite{KliDorPep-1980,TsuStoGos-1982}, new experiments
revealed resistivity oscillations~\cite{ZudSimRen-2001,YeEngTsu-2001} and
zero-resistance states~\cite{ManSmeKli-2002,ZudPfeWes-2003}, if a 2D electron
gas formed is GaAs/AlGaAs heterostructures is exposed to microwave (MW)
radiation. These oscillations are controlled by the ratio of the radiation
frequency, $\omega $, to the cyclotron frequency, $\omega _{c}$. The
zero-resistance states (ZRS) are assumed~\cite{AndAleMil-2003} to be caused
by instability of an electron system with absolute negative conductivity, $\sigma _{xx}<0$, regardless of the
actual mechanism of MW-induced resistance oscillations (MIRO) which
is still under debate (for a review, see Ref.~\onlinecite{DmiMirPol-2012}).

Among different theoretical mechanisms proposed for the explanation of MIRO,
here we would like to highlight the displacement~\cite{Ryz-1970,DurSacGir-2003}
and inelastic~\cite{DmiMirPol-2003} models. The displacement mechanism is based
on a peculiarity of orbit center migration ($%
X\rightarrow X^{\prime }$) when an electron absorbs a photon and
simultaneously is scattered off impurities. The authors of the inelastic mechanism
noticed that photon-assisted scattering affects the distribution function of electrons $%
f\left( \varepsilon \right) $ in such a way that it acquires a
nonequilibrium oscillating correction (a sort of population inversion) whose
derivative leads to a sign-changing contribution to $\sigma _{xx}$.

MW-induced magnetoconductivity oscillations similar to MIRO and even ZRS
were observed in a nondegenerate 2D electron gas formed on the free surface
of liquid helium~\cite{KonKon-2009,KonKon-2010}. The important distinction of
these new oscillations is that they are observed only if the excitation
energy of the second surface subband $\Delta _{2,1}\equiv \Delta _{2}-\Delta
_{1}$ is tuned to the resonance with the MW field ($\Delta _{2,1}=\hbar
\omega $) by varying the pressing electric field (a sort of Stark effect in
the 1D potential well formed at the surface). It should be noted also that the shape
of these oscillations strikingly differs from the usual shape of magnetointersubband
oscillations described theoretically~\cite{RaiSha-1994} and observed~\cite{SanHolHar-1998}
for semiconductor heterostructures under conditions that two subbands are occupied. Instead of
simple peaks of $\sigma _{xx}$ expected at the conditions of alignment of Landau levels
belonging to different subbands, the shape of MIRO observed in a 2D electron gas on
liquid helium represents rather a derivative of peaks.

The oscillations reported for electrons on liquid helium were
explained~\cite{Mon-2011a,Mon-2011b,Mon-2012} by a nonequilibrium population of the
excited subband which triggers quasi-elastic intersubband scattering of
electrons with the same peculiarity of orbit center migration as that
noticed in the displacement model. Thus, the intersubband mechanism of
MIRO and ZRS has something in common with the both
displacement and inelastic mechanisms though it does not use the concept of
photon-assisted scattering which is important for these two models.
Extensive studies of MIRO in a nondegenerate 2D electron gas on liquid helium
have revealed a number of remarkable effects associated with the ZRS regime:
in-plane redistribution of electrons~\cite{KonCheKon-2012}, self-generated
audio-frequency oscillations~\cite{KonWatKon-2013}, and incompressible states%
~\cite{CheWatNas-2015}. An explanation of these novel observations is based
on the concept of electron density domains~\cite{Mon-2016}: regions of
different densities appear to eliminate the regime of negative conductivity.

It should be noted also that even the delicate theoretical predictions reported for the
intersubband mechanism of MIRO~\cite{Mon-2012} which concern
the effect of Coulomb interaction on conductivity extrema were clearly observed
in the experiment~\cite{KonMonKon-2013}. Still, this mechanism of MIRO was described
only for a nondegenerate multisubband electron system using an important
simplification: $f\left( \varepsilon \right) \propto \exp \left(
-\varepsilon /T_{e}\right) $, where $\varepsilon $ is the in-plane energy,
and $T_{e}$ is the electron temperature. It is not clear how the Pauli
exclusion principle affects this mechanism; and the theory does not indicate
in what respect the results obtained for electrons on liquid helium can be applied to a
degenerate 2D electron system similar to those investigated in
semiconductor structures.

In this work we develop a theory of magnetoconductivity
oscillations in a degenerate 2D electron gas which are induced by nonequilibrium population
of excited subbands. We introduce a new definition of the extended dynamic structure
factor of a multisubband 2D electron system which incorporates the
concept of quasi-Fermi levels ($\mathit{imref}$) and describes the contribution
of elastic inter-subband scattering to the momentum relaxation rate under
conditions that electron distribution is strongly displaced from
equilibrium and cannot be attributed to simple heating of electrons. We demonstrate that
nonequilibrium populations of excited subbands can lead to
magnetointersubband oscillations whose shape differs from the shape of usual oscillations
caused by the equilibrium population of the second subband and the
alignment of staircases of Landau levels~\cite{RaiSha-1994}. This induces important changes
in quantum magnetotransport of a degenerate 2D electron system and can even lead to
negative linear response conductivity.

\section{Magnetotransport in multisubband 2D systems}

Electrons formed on the free surface of liquid helium have a rather low density
$n_{e}\lesssim 2 \times 10^{9}\, \mathrm{cm}^{-2}$, therefore at temperatures which are
comparable with the Fermi temperature they are already localized in sites of the Wigner
lattice~\cite{GriAda-1979}. Above the Wigner solid transition temperature this system
can be considered as a nondegenerate Coulomb liquid where the Pauli exclusion principle
is unimportant. Electrons on a liquid helium film represent a remarkable exception: for
a special arrangement of various substrates~\cite{PeePla-1983} they can form a 2D Fermion
system even at $T=0$.

Electrons in semiconductor structures usually have the effective mass which is much smaller than
the free electron mass. Therefore, at low temperatures these electrons can be described as a 2D Fermi gas.
A 2D electron system formed in a semiconductor device can have more
than one subband~\cite{AndFowSte-1982,Bas-1992,Dav-1998}. There is a number of experiments
demonstrated importance of intersubband scattering for electron transport
in a 2D system~\cite{StoGosWie-1982,SanHolHar-1998}. These results represent properties
of an equilibrium system, when the gate potential and the Fermi level position in a GaAs/AlGaAs
heterostructure provide the second subband occupancy. There is also a possibility of changing
carrier density by illuminating samples with light due to
electron-hole pair generation~\cite{HarLacFox-1987}. In this work, we shall focus
on magnetotransport properties of a 2D electron system under conditions that electron populations of excited
subbands deviate substantially from equilibrium and cannot be described by a single chemical potential.

The energy spectrum of a multisubband 2D electron system in crossed
magnetic ($B$) and electric ($E_{\Vert }$) fields is described by three
quantum numbers ($l$, $n$, and $X$; here we shall ignore the spin variable):%
\begin{equation}
\mathcal{E}_{l,n,X}=\Delta _{l}+\varepsilon _{n}+eE_{\Vert }X\text{ ,}
\label{eq1}
\end{equation}%
where $\Delta _{l}$ is the subband energy ($l=1,2,...$), $X$ is the
coordinate of the center of the cyclotron motion, $\varepsilon _{n}$ is
the usual Landau spectrum %
\begin{equation}
\varepsilon _{n}=\hbar \omega _{c}\left( n+1/2\right) ,  \label{eq2}
\end{equation}%
($n=0,1,...$), and $\omega _{c}=eB/m_{e}c$ is the cyclotron frequency.
In the center-of-mass reference frame moving with regard to the laboratory frame with the
drift velocity $\mathbf{u}_{d}$, the electric field $E_{\Vert }^{\prime
}\rightarrow 0$ and the in-plane electron motion is described by the pure
Landau spectrum of Eq.~(\ref{eq2}). The degeneracy of each Landau level is
given by $S_{\mathrm{A}}/2\pi \ell _{B}^{2}$, where $\ell _{B}=\sqrt{\hbar
c/eB}$ is the radius of the cyclotron orbit at $n=0$, and $S_{\mathrm{A}}$
is the surface area.

\begin{figure}[tbp]
\begin{center}
\includegraphics[width=9.0cm]{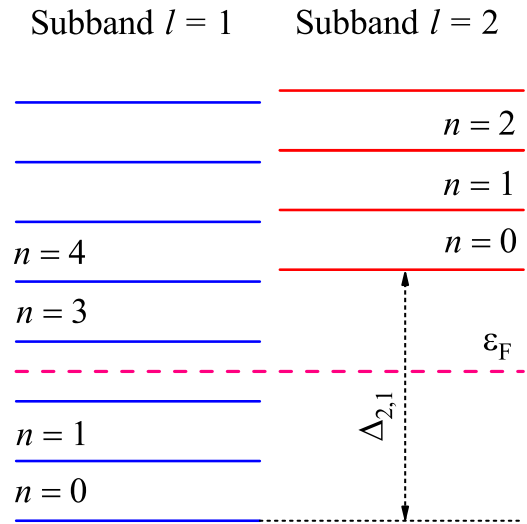}
\end{center}
\caption{ Schematic illustration of a two-subband 2D electron system in a magnetic field.
The energy spectrum of the ground (blue) and the first excited (red) subbands
represents a staircase of Landau levels. The position of the Fermi-level at equilibrium
is shown by the pink horizontal line.
} \label{f1}
\end{figure}

The schematic view of Landau levels of a two-subband system is shown in Fig.%
~\ref{f1}. The Landau levels of the excited subband are up-shifted by $\Delta
_{2,1}\equiv \Delta _{2}-\Delta _{1}$ as compared to respective levels of
the ground subband. In contrast with the model considered
previously~\cite{RaiSha-1994}, the equilibrium Fermi energy
$\varepsilon _{\mathrm{F}}$ is assumed
to be smaller than the intersubband excitation energy $\hbar \omega
_{2,1}=\Delta _{2,1}$ (here $\omega _{2,1}$ is the excitation frequency). It
is obvious that at certain magnetic fields defined by the condition $\omega
_{2,1}/\omega _{c}=m$ (here $m=1,2,...$) Landau levels of the excited subband becomes
completely aligned with high enough Landau levels of the ground subband
which triggers elastic intersubband scattering.

At strong magnetic fields directed perpendicular to the
electron layer, magnetotransport of a 2D electron gas is well described%
~\cite{AndUem-1974} by the center-migration theory~\cite{KubHasHas-1959,KuMiyHas-1965}, if
the collision broadening of Landau levels is taken into account. For
semiconductor electrons, there are two scattering mechanisms important at
low temperatures: Coulomb scattering from charged centers and surface
roughness scattering~\cite{AndFowSte-1982}. Both of them represent
essentially elastic scattering process. Each experimental realization of a
2D electron system has its own specific nature of scatterers.
The details of this nature are not important
for the effect considering in this work, and they can be incorporated in the
theory by changing the matrix elements of electron scattering.
As we shall see, the important parameters of the theory are the Landau level broadening
and the momentum collision rate at zero magnetic field.
Therefore, here we shall model the scatterers by artificial heavy atoms
interacting with electrons by an arbitrary potential $V_{\mathrm{int}}\left(
\left\vert \mathbf{R}_{e}-\mathbf{R}_{a}\right\vert \right) $ (here $\mathbf{%
R}_{e}$ and $\mathbf{R}_{a}$ are radius vectors of an electron and an atom
respectively).

In the model considering here, the interaction Hamiltonian can be represented
in terms of creation ($a_{\mathbf{K}}^{\dag }$) and destruction ($a_{\mathbf{%
K}}$) operators of atoms as%
\[
H_{\mathrm{int}} =\frac{1}{\Omega _{\mathrm{v}}}%
\sum_{e}\sum_{\mathbf{K},\mathbf{K}^{\prime }}\exp \left[ -i\left( \mathbf{K}%
^{\prime }-\mathbf{K}\right) \mathbf{R}_{e}\right]\times
\]
\begin{equation}
\times V_{\left\vert \mathbf{K}%
^{\prime }-\mathbf{K}\right\vert }a_{\mathbf{K}^{\prime }}^{\dag }a_{\mathbf{%
K}}\text{ },  \label{eq3}
\end{equation}%
where $\Omega _{\mathrm{v}}\equiv S_{\mathrm{A}}L_{z}$ is the volume
containing these atoms, $\mathbf{K}$ represents a 3D wave vector of an atom,
and $V_{\left\vert \mathbf{K}^{\prime }\mathbf{-K}\right\vert }$ is a
Fourier-transform of the potential $V_{\mathrm{int}}\left( R\right) $.
For the effective potential $V_{a}\delta \left( \mathbf{R}_{e}-\mathbf{R}_{a}\right) $,
conventionally describing interaction with short-range scatterers, $V_{Q}=V_{a}$. Static
defects resulting in elastic electron scattering are described by the
limiting case $M_{a}\rightarrow \infty $ (here $M_{a}$ is the mass of an
artificial atom). Surface defects can be modeled by a 2D layer of artificial atoms.
Similar modeling can be considered for a description of remote scatterers.

In the case of a nondegenerate 2D electron gas, the problem of finding the nonequilibrium
magnetoconductivity $\sigma _{xx}$ can be equally well solved by considering
the momentum exchange at a collision in the laboratory~\cite{Mon-2011a,Mon-2012,Mon-2013}
or in the center-of-mass\cite{Mon-2011b}
reference frames. For nondegenerate electrons, a great simplification
appears because $\left[ 1-f\left( \varepsilon _{n^{\prime },X^{\prime
}}\right) \right] \simeq 1$, and the quantity to be averaged in the
laboratory frame is independent of $X$. This allows one to restrict the averaging
procedure to the Landau level index $n$ only, assuming the distribution
function $f\left( \varepsilon _{n}\right) \propto \exp \left( -\varepsilon
_{n}/T_{e}\right) $ with an effective temperature $T_{e}$.

Magnetoconductivity $\sigma _{xx}$ of a degenerate 2D electron system
can be found from the average friction force $\mathbf{F}_{\mathrm{fr}}$
acting on electrons due to interaction with scatterers (the momentum balance
method~\cite{LeiTin-1984,CaiLeiTin-1985,MonTesWyd-2002,MonKon-book})
or using a direct expression for the current $%
j_{x}$ and calculating probabilities of electron scattering from $X$ to $%
X^{\prime }$ (a version of the Titeica's method~\cite{Tit-1935}). In order to
avoid complications with the field term $eE_{\Vert }X$ in the energy
spectrum of degenerate electrons, it is convenient to consider
scattering processes in the center-of-mass reference frame moving with the
drift velocity $\mathbf{u}_{\mathrm{d}}$ with regard to the laboratory
reference frame. In this moving frame, the driving electric field $%
E_{\Vert }^{\prime }$ is zero~\cite{MonKon-book}, and the electron spectrum
coincides with the Landau spectrum $\varepsilon _{n}$. It is important that
the momentum exchange at a collision $\mathbf{Q}\equiv \mathbf{K}^{\prime }%
\mathbf{-K}$ in the center-of-mass frame is the same as in the laboratory
frame because of the linear relationship between a momentum and the
respective velocity. At the same time, one have to keep in mind that in the
center-of-mass reference frame the energy exchange at an elastic collision
acquires a Doppler shift correction,~\cite{MonKon-book}%
\begin{equation}
E_{\mathbf{K}^{\prime }}^{\left( a\right) }-E_{\mathbf{K}}^{\left( a\right)
}=-\hbar \mathbf{Q}\cdot \mathbf{u}_{\mathrm{d}}\equiv -\hbar \mathbf{q}%
\cdot \mathbf{u}_{\mathrm{d}},  \label{eq4}
\end{equation}%
due to the quadratic dependence of the energy of an atom on its velocity.
Here $E_{\mathbf{K}}^{\left( a\right) }=\hbar ^{2}K^{2}/2M_{a}$ and we used
the notation $\mathbf{Q=}\left\{ \mathbf{q},\kappa \right\} $ with $\mathbf{q%
}$ and $\kappa $ standing for the in-plane and vertical components
respectively. It is quite obvious that scattering probabilities should not
depend on a choice of an inertial reference frame. Physically, the
correction of Eq.~(\ref{eq4}) is equivalent to the energy exchange for the
electron spectrum considered in the laboratory frame $eE_{\Vert }\left(
X^{\prime }-X\right) =\hbar q_{y}V_{\mathrm{H}}$, here we have taken into
account that $X^{\prime }-X=q_{y}\ell _{B}^{2}$ due to the momentum
conservation and used the notation $V_{\mathrm{H}}=cE_{\Vert }/B$ for the
Hall velocity ($u_{\mathrm{d}}^{\left( y\right) }\simeq -V_{\mathrm{H}}$).

The momentum balance approach\cite{MonTesWyd-2002, MonKon-book} allows
obtaining the effective collision frequency of electrons $\nu _{\mathrm{eff}}
$ from the kinetic friction acting on the whole electron system $\mathbf{F}_{%
\mathrm{fr}}$. In the linear transport regime, $\mathbf{F}_{\mathrm{fr}}$ is
proportional to $\mathbf{u}_{\mathrm{d}}$, and conventionally it can be
written as~\cite{PetSchMon-1994} $\mathbf{F}_{\mathrm{fr}}=-N_{e}m_{e}\nu _{\mathrm{eff}}\mathbf{u}%
_{\mathrm{d}}$, where the proportionality factor $\nu _{\mathrm{eff}}$
defines electron magnetoconductivity%
\begin{equation}
\sigma _{xx}\simeq \frac{e^{2}n_{e}\nu _{\mathrm{eff}}}{m_{e}\omega _{c}^{2}}%
\text{ },  \label{eq5}
\end{equation}%
and $n_{e}=N_{e}/S_{\mathrm{A}}$ is electron density.

The simplest way of
obtaining $\nu _{\mathrm{eff}}$ is to consider the momentum balance along
the $y$-axis, $F_{\mathrm{fr}}^{\left( y\right) }=-N_{e}m_{e}\nu _{\mathrm{eff%
}}u_{\mathrm{d}}^{\left( y\right) }$. Assuming $u_{\mathrm{d}}^{\left(
y\right) }\simeq -V_{\mathrm{H}}$ and using the Born approximation for scattering
probabilities in the center-of-mass frame, one can find
\begin{equation}
F_{\mathrm{fr}}^{\left( y\right) }\left( V_{\mathrm{H}}\right) =-N_{e}\sum_{%
\mathbf{q}}\hbar q_{y}\bar{W}_{\mathbf{q}}\left( V_{\mathrm{H}}\right) ,
\label{eq6}
\end{equation}%
where
\[
\bar{W}_{\mathbf{q}}\left( V_{\mathrm{H}}\right) =\frac{2\pi n_{a}^{\left( 3%
\mathrm{D}\right) }}{\eta \hbar S_{\mathrm{A}}}\sum_{l,l^{\prime
},n,n^{\prime }}f_{l}\left( \varepsilon _{n}\right) \left[ 1-f_{l^{\prime
}}\left( \varepsilon _{n^{\prime }}\right) \right] \times
\]
\begin{equation}
\times I_{n,n^{\prime }}^{2}\left( x_{q}\right) U_{l^{\prime },l}^{2}
\left( q\right) \delta \left( \varepsilon _{n^{\prime }}-\varepsilon _{n}+
\Delta _{l^{\prime },l}+\hbar q_{y}V_{\mathrm{H}}\right)   \label{eq7}
\end{equation}%
is the probability of electron scattering with the in-plane momentum
exchange equal $\hbar \mathbf{q}$, and $\Delta _{l^{\prime },l}=\Delta _{l^{\prime }}-\Delta _{l}$.
Here we have used the following
notations: $n_{a}^{\left( 3\mathrm{D}\right) }$ is the density of
scatterers, $\eta =2\pi \ell _{B}^{2}n_{e}$ is the filling factor, $%
f_{l}\left( \varepsilon _{n}\right) $ is the electron distribution function,
the functions $U_{l^{\prime },l}^{2}\left( q\right) $ and $I_{n,n^{\prime
}}^{2}\left( x_{q}\right) $ are defined by matrix elements of the
interaction Hamiltonian
\begin{equation}
U_{l^{\prime },l}^{2}\left( q\right) =\frac{1}{L_{z}}\sum_{\kappa }V_{\sqrt{%
q^{2}+\kappa ^{2}}}^{2}\left\vert \left( e^{-i\kappa z}\right) _{l^{\prime
},l}\right\vert ^{2},  \label{eq8}
\end{equation}%
\begin{equation}
\left\vert \left( e^{-i\mathbf{q}\cdot \mathbf{r}_{e}}\right) _{n^{\prime
},X^{\prime };n,X}\right\vert ^{2}=\delta _{X,X^{\prime }-\ell
_{B}^{2}q_{y}}I_{n,n^{\prime }}^{2}\left( x_{q}\right) ,  \label{eq9}
\end{equation}%
\[
I_{n,n^{\prime }}^{2}(x)=\frac{[\min (n,n^{\prime })]!}{[\max (n,n^{\prime
})]!}x^{|n-n^{\prime }|}e^{-x}\left[ L_{\min (n,n^{\prime })}^{|n-n^{\prime
}|}(x)\right] ^{2},
\]%
$x_{q}=q^{2}\ell _{B}^{2}/2$, and $L_{n}^{m}(x)$ are the associated
Laguerre polynomials. When obtaining Eq.~(\ref{eq7}), we used the
advantages of describing scattering probabilities in the moving frame - the
summations over indexes $X$, $X^{\prime }$ and $\mathbf{K}$ are trivial
leading to the factors $n_{B}=1/2\pi \ell _{B}^{2}$ and $n_{a}^{\left( 3%
\mathrm{D}\right) }$.

Comparing the right side of Eq.~(\ref{eq6}) with the result expected for the
linear regime $N_{e}m_{e}\nu _{\mathrm{eff}}V_{\mathrm{H}}$, one can find that%
\begin{equation}
\nu _{\mathrm{eff}}=-\frac{1}{m_{e}V_{\mathrm{H}}}\sum_{\mathbf{q}}\hbar
q_{y}\bar{W}_{\mathbf{q}}\left( V_{\mathrm{H}}\right) .  \label{eq10}
\end{equation}%
When expanding $\bar{W}_{\mathbf{q}}\left( V_{\mathrm{H}}\right) $ in $V_{%
\mathrm{H}}$, we can consider only the linear term $\bar{W}_{\mathbf{q}%
}^{\prime }\left( 0\right) V_{\mathrm{H}}$ [here the 'prime' denotes the
differentiation] because $\bar{W}_{\mathbf{q}}\left( 0\right) $ depends only on
the absolute value of $\mathbf{q}$ and, therefore, gives zero contribution
into $\nu _{\mathrm{eff}}$.

It is instructive to note that the same result for $\nu _{\mathrm{eff}}$ and
$\sigma _{xx}$ can be found from the direct expression for the electron
current along $x$-direction (this method was also used~\cite{Mon-2017,MonKon-2019}
for describing a nondegenerate electron system):%
\begin{equation}
 j_{x}=-en_{e}\sum_{\mathbf{q}}\left( X^{\prime }-X\right) _{\mathbf{q}}\bar{W%
}_{\mathbf{q}}\left( V_{\mathrm{H}}\right) ,
\label{eq11}
\end{equation}%
where we have to use the relationship $\left( X^{\prime }-X\right) _{\mathbf{q}%
}=\ell _{B}^{2}q_{y}$ which follows from matrix elements of Eq.~(\ref{eq9}).
The Eq.~(\ref{eq11}) and the definition of $\sigma _{xx}$ obviously yield the
expression for $\nu _{\mathrm{eff}}$ given in Eq.~(\ref{eq10}).

To obtain a finite magnetoconductivity in the treatment presented above, one
have to include higher approximations by incorporating the collision
broadening of Landau levels $\Gamma _{l,n}$ (the broadening of electron
density of states). Following the ideas of the
center migration theory~\cite{KubHasHas-1959} and the self-consistent Born
approximation (SCBA)\cite{AndUem-1974}, in the right side of Eq.~(\ref{eq7}) we shall
insert $\int d\varepsilon \int d\varepsilon ^{\prime }\delta _{l}\left(
\varepsilon -\varepsilon _{n}\right) \delta _{l^{\prime }}\left( \varepsilon
^{\prime }-\varepsilon _{n^{\prime }}\right) $; the subscripts $l$ and
$l^{\prime }$ in the respective delta-functions just mark the subband where
the level density belongs.
Then, assuming the replacement $\delta _{l}\left( \varepsilon
-\varepsilon _{n}\right) \rightarrow -\frac{1}{\pi \hbar }\mathrm{Im}%
G_{l,n}\left( \varepsilon \right) $ [here $G_{l,n}\left( \varepsilon \right)
$ is the single-electron Green's function], the average
probability of scattering with the momentum exchange $\hbar \mathbf{q}$
can be represented in the following form:
\begin{equation}
\bar{W}_{\mathbf{q}}\left( V_{\mathrm{H}}\right) =\frac{n_{a}^{\left( 3%
\mathrm{D}\right) }}{S_{\mathrm{A}}\hbar ^{2}}\sum_{l,l^{\prime
}}U_{l^{\prime },l}^{2}\left( q\right) D_{l,l^{\prime }}\left( q,\omega
_{l,l^{\prime }}-q_{y}V_{\mathrm{H}}\right) ,  \label{eq12}
\end{equation}%
where $\omega _{l,l^{\prime }}=\Delta _{l,l^{\prime }}/\hbar $, and
\[
D_{l,l^{\prime }}\left( q,\Omega \right) =\frac{2}{\pi \hbar \eta }\int
d\varepsilon f_{l}\left( \varepsilon \right) \left[ 1-f_{l^{\prime }}\left(
\varepsilon +\hbar \Omega \right) \right] \times
\]
\begin{equation}
\times \sum_{n,n^{\prime }}I_{n,n^{\prime
}}^{2}\left( x_{q}\right) \mathrm{Im}G_{l,n}\left( \varepsilon \right) \mathrm{Im%
}G_{l^{\prime },n^{\prime }}\left( \varepsilon +\hbar \Omega \right) \text{ }
\label{eq13}
\end{equation}%
is a new generalization of the dynamic structure factor (DSF) of a
multisubband 2D electron system. Expanding $\bar{W}_{\mathbf{q}}$ in $%
q_{y}V_{\mathrm{H}}$ yields%
\begin{equation}
\nu _{\mathrm{eff}}=\frac{n_{a}^{\left( 3\mathrm{D}\right) }}{m_{e}\hbar
S_{A}}\sum_{\mathbf{q}}\sum_{l,l^{\prime }}q_{y}^{2}U_{l^{\prime
},l}^{2}\left( q\right) D_{l,l^{\prime }}^{\prime }\left( q,\omega
_{l,l^{\prime }}\right) \text{ .}  \label{eq14}
\end{equation}%
Thus, the effective collision frequency of a multisubband 2D electron system
is proportional to the derivative of the extended DSF $D_{l,l^{\prime }}^{\prime
}\left( q,\omega _{l,l^{\prime }}\right) $ with respect to frequency.

There are two important approximations for the Landau level density of
states. The SCBA theory of Ando and Uemura yields the semi-elliptical shape
of the density of states~\cite{AndUem-1974}%
\begin{equation}
-\mathrm{Im}G_{n}\left( \varepsilon \right) =\frac{2\hbar }{\Gamma _{n}}\sqrt{%
1-\frac{\left( \varepsilon -\varepsilon _{n}\right) ^{2}}{\Gamma _{n}^{2}}},
\label{eq15}
\end{equation}%
where $\Gamma _{n}$ is the broadening parameter. In the case of short-range
scatterers, $\Gamma _{n}$ is independent of Landau number $\Gamma
_{n}=\Gamma $ with~\cite{AndUem-1974}%
\begin{equation}
\Gamma =\sqrt{\frac{2}{\pi }\hbar \omega _{c}\nu _{0}}\text{ },  \label{eq16}
\end{equation}%
where $\nu _{0}$ is the electron relaxation rate obtained for $B=0$. The
cumulant expansion method~\cite{Ger-1976} yields the Gaussian shape of Landau
levels%
\begin{equation}
-\mathrm{Im}G_{n}\left( \varepsilon \right) =\frac{\sqrt{2\pi }\hbar }{\Gamma
_{n}}\exp \left[ -\frac{2\left( \varepsilon -\varepsilon _{n}\right) ^{2}}{%
\Gamma _{n}^{2}}\right] ,  \label{eq17}
\end{equation}%
which does not have the sharp cutoff of the density of states. Generally,
the level shape is a kind of mixture of elliptical and Gaussian
forms~\cite{And-1974}, and the shape of the lowest level is close to a Gaussian.

In the case of equilibrium Fermi-distribution, $D_{l,l^{\prime }}\left(
q,\Omega \right) $ has very useful properties which simplify significantly
evaluation of $\nu _{\mathrm{eff}}$ and $\sigma _{xx}$. For example,
consider only the contribution from intrasubband scattering processes ($%
l^{\prime }=l$). Then, $D_{l,l}\left( q,\Omega \right) $ coincides with the
conventional DSF of a 2D electron system which satisfies the condition%
\begin{equation}
D_{l,l}\left( q,-\Omega \right) =\mathrm{e}^{-\hbar \Omega
/T_{e}}D_{l,l}\left( q,\Omega \right)   \label{eq18}
\end{equation}%
The derivative of this relationship gives $D_{l,l}^{\prime }\left(
q,0\right) =\frac{\hbar }{2T_{e}}D_{l,l}\left( q,0\right) $ and the linear
(in $q_{y}V_{\mathrm{H}}$) term of Eq.~(\ref{eq12}) can be rewritten as
\begin{equation}
\delta \bar{W}_{\mathbf{q}}\simeq -q_{y}V_{\mathrm{H}}\frac{\hbar }{2T_{e}}%
\bar{W}_{\mathbf{q}}\left( 0\right) \text{ ,}  \label{eq19}
\end{equation}%
which allows representing $\sigma _{xx}$ in terms of the equilibrium
probability $\bar{W}_{\mathbf{q}}\left( 0\right) $:%
\begin{equation}
\sigma _{xx}\simeq \frac{e^{2}n_{e}}{2T_{e}}\sum_{\mathbf{q}}\left(
X^{\prime }-X\right) _{\mathbf{q}}^{2}\bar{W}_{\mathbf{q}}\left( 0\right)
\text{ }.  \label{eq20}
\end{equation}%
This equation coincides with the well-known result obtained
previously~\cite{KubHasHas-1959,AdaHol-1959}, and it is similar to the Einstein relation
between the conductivity and the diffusion coefficient.

For the ground subband and the semi-elliptic shape of Landau levels [Eq.~(\ref%
{eq15})] induced by short-range scatterers, Eq.~(\ref{eq20}) transforms into
the result of Ando and Uemura which indicates that the conductivity peak
value $\left( \sigma _{xx}\right) _{\mathrm{\max }}=\frac{e^{2}}{\pi
^{2}\hbar }\left( n+1/2\right) $ depends only on the Landau level index $n$
and the natural constants~\cite{AndUem-1974}. These "checkpoints" of
equilibrium transport regime, encourage us to use Eq.~(\ref{eq14}) for
describing magnetotransport in nonequilibrium multisubband 2D electron
systems.

For a nonequilibrium filling of 2D subbands, the extended DSF $%
D_{l,l^{\prime }}\left( q,\Omega \right) $ generally has no a relationship
similar to Eq.~(\ref{eq18}). Only describing nondegenerate electrons and assuming $%
f_{l}\left( \varepsilon \right) \propto N_{l}\exp \left( -\varepsilon
/T_{e}\right) $ it was possible to introduce~\cite{Mon-2011b,Mon-2012} a
version of the DSF $S_{l,l^{\prime }}\left( q,\Omega \right) $ which had an
important property resembling Eq.~(\ref{eq18}), in spite of the fact that
the occupation of subbands was not equilibrium. Unfortunately, this version of
the extended DSF appears to be useless for degenerate electrons.
The new definition of the extended DSF $%
D_{l,l^{\prime }}\left( q,\Omega \right) $ given in Eq.~(\ref{eq13})
transforms into $\bar{n}_{l}S_{l,l^{\prime }}\left( q,\Omega \right) $ if
the electron system can be considered as a nondegenerate gas [here $\bar{n}_{l}=N_{l}/N_{e}$ is
the fractional occupancy of a subband].

\section{Quasi-Fermi level approximation}

Generally, it is very difficult to find $f_{l}\left( \varepsilon \right) $
if a system is displaced from equilibrium. Therefore, in solid state physics
it is quite common to use the concept of a quasi-Fermi level or $\mathit{imref}$%
. In the following, we assume that displacement from equilibrium is such
that electron populations can no longer be described by a single chemical
potential (or a Fermi level), nevertheless it is possible to describe it
introducing separate chemical potentials (quasi-Fermi levels) for each
subband:%
\begin{equation}
f_{l}\left( \varepsilon \right) =\frac{1}{\mathrm{e}^{\left( \varepsilon
+\Delta _{l,1}-\mu _{l}\right) /T_{e}}+1}\equiv f_{\mathrm{F}}\left( \varepsilon
+\Delta _{l,1}-\delta \mu _{l}\right) \text{ },  \label{eq21}
\end{equation}%
where $\delta \mu _{l}=\mu _{l}-\mu $. The chemical potentials $\mu _{l}$
are measured from the bottom of the ground subband, while the zero of Landau
energy $\varepsilon $ is taken at the bottom of each subband. In most cases,
it is sufficient to consider only two subbands (the ground subband and the
first excited subband), when electron populations of higher subbands can be
neglected. The form of Eq.~(\ref{eq21}) is quite accurate if
electron-electron collisions are more important for intrasubband
redistribution than for intersubband decay rates. Anyway, this form of $%
f_{l}\left( \varepsilon \right) $ is very useful because it allows obtaining
$\sigma _{xx}$ in an analytical form for nonequilibrium populations of
electron subbands.

One can also introduce different electron temperatures
for each subband ($T_{l,e}$), still we shall assume that $T_{l,e}=T_{l^{\prime },e}= T_{e}$
because in-plane energy relaxation between different subbands is governed by
electron-electron collisions (electron spacing is usually much larger than the
average distance between nearest subbands), whose rate is quite high for 2D electron
systems~\cite{AndFowSte-1982}. Regarding possible heating of electrons ($T_{e}>T$), we assume
that $T_{e}$ is still much lower than the quasi-Fermi energies. The opposite limiting case
(nondegenerate electrons) was described in Refs.~\onlinecite{Mon-2011b,Mon-2012}.
It should be noted also that MIRO observed in a 2D electron gas on liquid helium
are quite well described even by the approximation $T_{e}=T$ in spite of a substantial
heating~\cite{KonMonKon-2013}.

Using the distribution function of Eq.~(\ref{eq21}) and the well-known identity%
\begin{equation}
f_{\mathrm{F}}\left( \varepsilon \right) \left[ 1-f_{\mathrm{F}}\left( \varepsilon ^{\prime }\right) %
\right] =\left[ f_{\mathrm{F}}\left( \varepsilon \right) -f_{\mathrm{F}}\left( \varepsilon ^{\prime
}\right) \right] \frac{1}{1-\mathrm{e}^{\left( \varepsilon -\varepsilon
^{\prime }\right) /T_{e}}}\text{ ,}  \label{eq22}
\end{equation}%
it is possible to establish the following relationship for the
extended DSF~\cite{comment}%
\begin{equation}
D_{l^{\prime },l}\left( q,-\Omega \right) =\mathrm{e}^{-\left( \hbar \Omega
+\Delta _{l^{\prime },l}-\mu _{l^{\prime },l}\right) /T_{e}}D_{l,l^{\prime
}}\left( q,\Omega \right) ,  \label{eq23}
\end{equation}%
where $\mu _{l^{\prime },l}=\mu _{l^{\prime }}-\mu _{l}$. For a single
subband ($l^{\prime }=l$), this property coincides with the property of the
usual DSF of a 2D electron gas given in Eq.~(\ref{eq18}).

When considering the contribution from intersubband scattering $\nu _{%
\mathrm{inter}}$ in Eq.~(\ref{eq14}), the property of Eq.~(\ref{eq23}) allows
us to transform derivatives of the DSF whose frequency argument is negative
into functions with a positive argument
\[
D_{l^{\prime },l}^{\prime }\left( q,-\Omega \right) =-\mathrm{e}^{-\left(
\hbar \Omega +\Delta _{l^{\prime },l}-\mu _{l^{\prime },l}\right)
/T_{e}}D_{l,l^{\prime }}^{\prime }\left( q,\Omega \right) +
\]
\begin{equation}
+\frac{\hbar }{T_{e}}\mathrm{e}^{-\left( \hbar \Omega
+\Delta _{l^{\prime },l}-\mu
_{l^{\prime },l}\right) /T_{e}}D_{l,l^{\prime }}\left( q,\Omega \right)
\text{ }  \label{eq24}
\end{equation}%
Thus, a substantial part of $D_{l,l^{\prime }}^{\prime
}\left( q,\omega _{l,l^{\prime }}\right) $ entering Eq.~(\ref{eq14}) can be
eliminated by reverse scattering processes due to the first term in the
right side of Eq.~(\ref{eq24}). Therefore, it is convenient to represent the
contribution of intersubband scattering to $\nu _{\mathrm{eff}}$ in the form
containing only positive frequency arguments ($l>l^{\prime }$). In this way, one
can obtain two kinds of contributions: a normal contribution proportional to $%
D_{l,l^{\prime }}\left( q,\omega _{l,l^{\prime }}\right) $, and an abnormal
(sign-changing) contribution proportional to the derivative $D_{l,l^{\prime
}}^{\prime }\left( q,\omega _{l,l^{\prime }}\right) $. To make a distinction
between these contributions, we shall use the following notations: $\nu _{\mathrm{inter%
}}=\nu _{\mathrm{N}}+\nu _{\mathrm{A}}$, where
\begin{equation}
\nu _{\mathrm{N}}=\frac{n_{a}^{\left( 3%
\mathrm{D}\right) }}{m_{e}\hbar S_{A}}\sum_{l>l^{\prime }}\sum_{%
\mathbf{q}}\frac{\hbar }{T_{e}}q_{y}^{2}U_{l^{\prime },l}^{2}\left( q\right)
\mathrm{e}^{-\mu _{l,l^{\prime }}/T_{e}}D_{l,l^{\prime }}\left( q,\omega
_{l,l^{\prime }}\right),  \label{eq25}
\end{equation}%
\[
\nu _{\mathrm{A}}=\frac{n_{a}^{\left( 3%
\mathrm{D}\right) }}{m_{e}\hbar S_{A}}\sum_{l>l^{\prime
}}\left( 1-\mathrm{e}^{-\mu _{l,l^{\prime }}/T_{e}}\right) \times
\]
\begin{equation}
\times \sum_{\mathbf{q}}q_{y}^{2}U_{l^{\prime },l}^{2}\left( q\right)
D_{l,l^{\prime }}^{\prime }\left( q,\omega _{l,l^{\prime }}\right)
\text{ }. \label{eq26}
\end{equation}%
The normal contribution $\nu _{\mathrm{N}}$ exists even under the
equilibrium condition ($\mu _{l,l^{\prime }}=0$), though at $\mu <\Delta
_{l,l^{\prime }}$ it is very small due to $f_{l}\left( \varepsilon \right) $
present in $D_{l,l^{\prime }}\left( q,\omega _{l,l^{\prime }}\right) $.
The abnormal terms $\nu _{\mathrm{A}}$ differ from zero only if electron
distribution is somehow displaced from equilibrium ($\mu _{l,l^{\prime }}>0$).

\begin{figure}[tbp]
\begin{center}
\includegraphics[width=10.0cm]{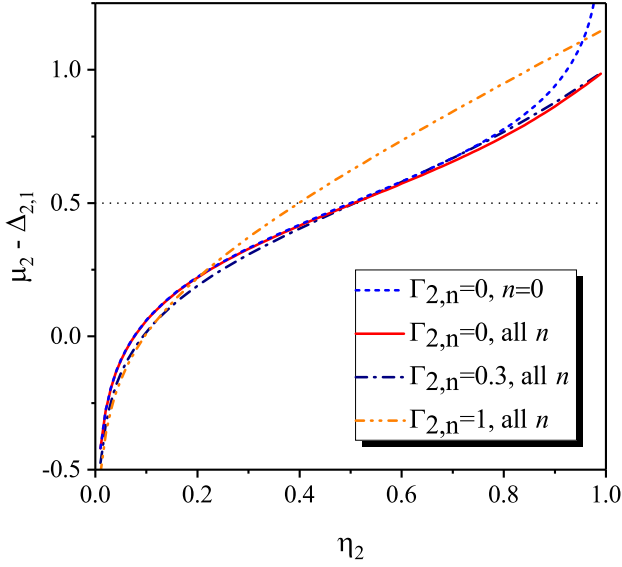}
\end{center}
\caption{ The quasi-chemical potential of the first excited subband
$\mu _{2}-\Delta _{2,1}$ (in units of $\hbar \omega _{c}$) versus the filling factor $\eta
_{2}=2\pi \ell _{B}^{2}N_{2}/S_{\mathrm{A}}$ calculated for different conditions which are
indicated in the figure legend. The level broadening $\Gamma _{2,n}$ is also shown in units
of $\hbar \omega _{c}$.
} \label{f2}
\end{figure}

When the first excited subband ($l=2$) has an extra electron population $%
\delta N_{2}$, one expects that the all these electrons will occupy the
lowest Landau level ($n=0$), if low temperatures ($T_{e}\ll \hbar \omega _{c}
$) are considered and the filling factor of the excited subband $\eta
_{2}=2\pi \ell _{B}^{2}N_{2}/S_{\mathrm{A}}<1$. Neglecting electron
populations at higher Landau levels and assuming that the level broadening
is small, one can find the quasi-Fermi level of the excited subband
\begin{equation}
\text{\ }\mu _{2}=\Delta _{2,1}+\varepsilon _{0}-T_{e}\ln \left( \frac{%
1-\eta _{2}}{\eta _{2}}\right) \text{ }.  \label{eq27}
\end{equation}%
In this equations, the last two terms represent the well-known high-field
approximation for the chemical potential~\cite{Mah-book-2000}.

The influence
of higher Landau levels and a finite broadening $\Gamma _{2,0}$ on $\mu
_{2}\left( \eta _{2}\right) $ is illustrated in Fig.~\ref{f2} for $\hbar
\omega _{c}/T_{e}=5$ (in this figure $\mu _{2}-\Delta _{2,1}$ and $\Gamma _{2,n}$ are
given in units of $\hbar \omega _{c}$). These results indicate that the
simple form of Eq.~(\ref{eq27}) describes the dependence $\mu _{2}\left( \eta
_{2}\right) -\Delta _{2,1}$ quite well if $\Gamma _{2,0}/\hbar \omega
_{c}\leq 0.3$. At $\Gamma _{2,0}/\hbar \omega _{c}=0.1$, it is even
difficult to see the difference between results of numerical calculations
(not shown in Fig.~\ref{f2}) and the approximation $\Gamma _{2,0}=0$
illustrated in the figure by the red line. For the strong broadening $\Gamma
_{2,0}/\hbar \omega _{c}=1$, the results of numerical calculations (orange
line) deviate substantially from the approximation given in Eq.~(\ref{eq27}),
if $\eta _{2}>0.2$. Under these conditions, the analytical form can be used
only for a qualitative analysis or simple estimations. It is important that
considering a 2D electron system with narrow Landau levels, the approximation
of Eq.~(\ref{eq27}) can be used even for substantial values of the filling
factor $\eta _{2}\leq 0.8$ which are quite sufficient for this
research. The accuracy of the high field approximation increases with
lowering temperature.

\begin{figure}[tbp]
\begin{center}
\includegraphics[width=10.0cm]{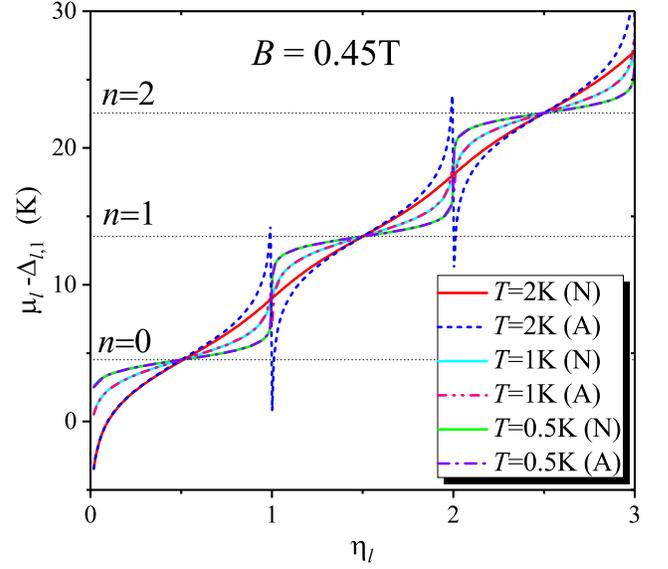}
\end{center}
\caption{The analytical (A) extension of the
quasi-chemical potential of the $l$-subband (dashed, dash-dotted and
dash-dot-doted lines) is compared with the results of numerical (N)
calculations for narrow Landau levels (solid lines). The wavy shape of the solid
lines increases with lowering temperature together with the accuracy of the
analytical approximation.
} \label{f3}
\end{figure}

For larger values of the filling factor $\eta _{2}>1$, one can find a simple
extension of the analytical form of Eq.~(\ref{eq27}) which can be used for
the ground subband as well. Therefore, in the following equation, we shall
use an arbitrary subband index ($l$):
\[
\mu _{l} -\Delta _{l,1}=\sum_{n=0}^{\infty }\left[
\varepsilon _{n}-T_{e}\ln \left( \frac{n+1-\eta _{l}}{\eta _{l}-n}\right) %
\right] \times
\]
\begin{equation}
\times \theta \left( n+1-\eta _{l}\right) \theta \left( \eta _{l}-n\right)
\text{ ,}  \label{eq28}
\end{equation}%
where $\theta \left( x\right) $ is the Heaviside step function, and $\eta
_{l}=2\pi \ell _{B}^{2}N_{l}/S_{\mathrm{A}}$. This solution is found
assuming that $f_{l}\left( \varepsilon \right) \simeq 1$ for $\varepsilon
\leq \varepsilon _{n-1}$ if $n<\eta _{l}<n+1$, therefore it is a low
temperature approximation. Fig.~\ref{f3} illustrates that at low temperatures
($T_{e}\leq 1\,\mathrm{K}$) numerical results shown by solid lines, are well
approximated by the periodic extension of the high field formula of Eq.~(\ref%
{eq27}) given in Eq.~(\ref{eq28}). Deviations of Eq.~(\ref{eq28}) from
the numerical result appear only in very narrow regions near the points
$\eta _{l}=1,2,...$. At high temperatures $T_{e}\gtrsim 0.2\,\hbar \omega
_{c}$, the deviations are strong because the numerical results shown by the red
line approach the semi-classical formula $\mu _{l}\left( \eta _{l}\right)
-\Delta _{l,1}\simeq 2\pi \hbar ^{2}N_{l}/m_{e}S_{\mathrm{A}}$.
In our numerical calculations (here and below), the ratio of the effective
electron mass to the free electron mass is fixed to the value 0.067 which is typical
for semiconductor heterostructures.

\begin{figure}[tbp]
\begin{center}
\includegraphics[width=10.0cm]{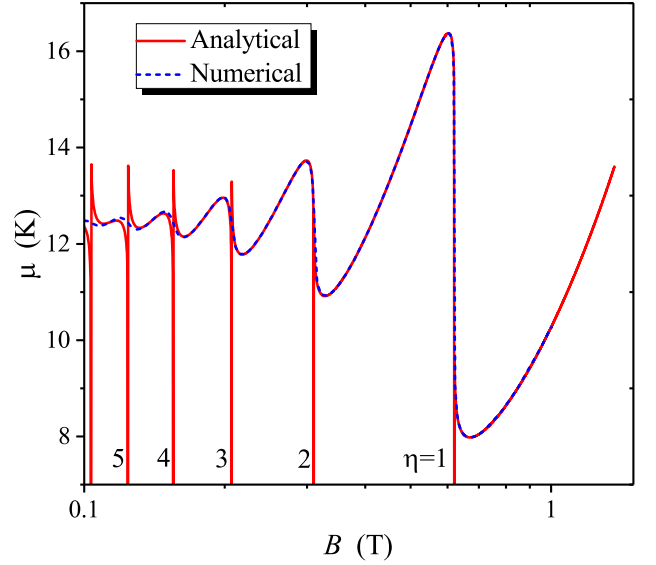}
\end{center}
\caption{Illustration of the efficiency of the analytical approximation given
in Eq.~(\ref{eq28}) for the description of oscillations of the chemical
potential (Fermi energy) as a function of $B$ under conditions that the
collision broadening is small: the analytical equation (solid red line),
and numerical calculations (dashed blue line). The singular points,
where the filling factor $\eta $ equals to an integer, are indicated.
} \label{f4}
\end{figure}

In Fig.~\ref{f3}, the filling factor $\eta _{l}$ was varied by changing
electron density $n_{l}=N_{l}/S_{\mathrm{A}}$, while the magnetic field was
fixed. It is remarkable that the simple analytical approximation given in
Eq.~(\ref{eq28}) can be used also for the description of the well-known
oscillations~\cite{ZawLas-1984} of the chemical potential $\mu \left( B\right) $ of a 2D
electron system with a fixed density and narrow Landau levels (here we omit
the subband index). This possibility is illustrated in Fig.~\ref{f4} for $%
n_{e}=1.5\times 10^{10}\,\mathrm{cm}^{-2}$ and $T=0.5\,\mathrm{K}$, assuming that
the broadening of Landau levels is small. One can see that the analytical
formula (red line) practically coincides with the results of numerical
calculations (blue line) in a wide range of magnetic fields with the
exception of the points where $\eta \left( B\right) $ is very close to an
integer ($1,2,...$) as indicated in Fig.~\ref{f4}.

\section{Results and discussion}

According to Eqs.~(\ref{eq25}) and (\ref{eq26}) the contribution from
intersubband scattering to the effective collision frequency as a function
of the magnetic field is determined by the extended DSF $D_{l,l^{\prime
}}\left( q,\Omega \right) $ and its derivative with respect to frequency $%
D_{l,l^{\prime }}^{\prime }\left( q,\Omega \right) $ near the special points
$\Omega = \omega _{l,l^{\prime }}\equiv \Delta _{l,l^{\prime }}/\hbar
>0$. Considering the two subband model ($l=2$ and $l^{\prime }=1$), in Eq.~(%
\ref{eq13}) which defines $D_{2,1}\left( q,\Omega \right) $
the factor $\left[ 1-f_{1}\left( \varepsilon +\hbar \Omega \right) \right] $
can be set to unity
because the distribution function of electrons occupying the ground subband
is very small at high energies: $f_{1}\left( \varepsilon +\Delta
_{2,1}\right) \ll 1$. The later inequality follows from the fact that the
respective quasi Fermi level $\mu _{1}\leq \mu $. For the regime of fixed
density, $\mu _{1}<\mu $ which is quite obvious according to Fig.~\ref{f3}.
In the the regime of fixed chemical potential, $\mu _{1}=\mu $ due to a
reservoir of electrons~\cite{Mah-book-2000}. Therefore, the nonequilibrium
DSF $D_{2,1}\left( q,\Omega \right) $ as a function of frequency is
determined mostly by the distribution of
electrons occupying the excited subband%
\begin{equation}
f_{2}\left( \varepsilon \right) =\left\{ \frac{1-\eta _{2}}{\eta _{2}}\exp
\left( \frac{\varepsilon -\varepsilon _{0}}{T_{e}}\right) +1\right\} ^{-1},
\label{eq29}
\end{equation}%
where we had used the approximation of Eq.~(\ref{eq27}) for \ $\mu _{2}$
assuming that $\eta _{2}\leq 0.8$. For larger $\eta _{2}$, we shall use the
extension of Eq.~(\ref{eq28}).

\begin{figure}[tbp]
\begin{center}
\includegraphics[width=10.0cm]{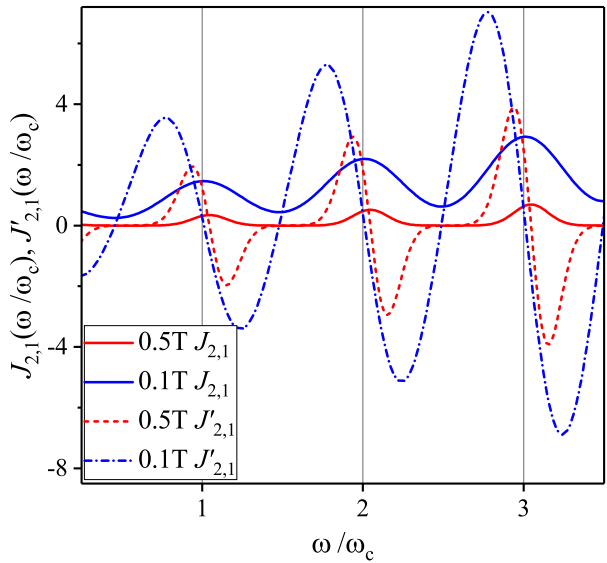}
\end{center}
\caption{The frequency dependencies of the dimensionless functions which define
the shape of magnetooscillations of $\nu _{\mathrm{N}}$ and $\nu _{\mathrm{A}}$
calculated for two values of the magnetic field shown in the figure legend:
$J_{2,1}\left( \omega /\omega _{c}\right) $ (solid lines) and $J^{\prime }_{2,1}\left( \omega /\omega _{c}\right)$
(dashed and dash-dotted lines).
} \label{f5}
\end{figure}

In the expression for the effective collision frequency $\nu _{\mathrm{eff}}$%
, the DSF is affected by integration over $q$. For short-range scatterers,
the respective integral can
be easily calculated because $\int x_{q}I_{n,n^{\prime
}}^{2}(x_{q})dx_{q}=\left( n+n^{\prime }+1\right) $. Therefore, it is
convenient to analyze the frequency dependence of the dimensionless function
\begin{equation}
J_{2,1}\left( \omega /\omega _{c}\right) =\frac{\eta \Gamma }{%
4\hbar }\int\limits_{0}^{\infty }D_{2,1}\left( q,\omega \right) x_{q}dx_{q}
\label{eq30}
\end{equation}%
instead of $D_{2,1}\left( q,\omega \right) $. Here, for simplicity reasons,
the collision broadening of Landau levels $\Gamma $ is assumed to be
independent of quantum numbers $n$ and $l$.
Employing the Gaussian shape of $\mathrm{Im}G_{l,n}\left( \varepsilon
\right) $ given in Eq.~(\ref{eq17}) yields
\[
J_{2,1} =\eta _{2}\sum_{n=1}^{\infty
}\left( n+1\right) \int\limits_{-\varepsilon _{0}/\Gamma }^{\infty }\frac{%
\exp \left( -2y^{2}\right) }{\left( 1-\eta _{2}\right) \exp \left( y\Gamma
/T_{e}\right) +\eta _{2}}\times
\]%
\begin{equation}
\times \exp \left\{ -2\left[ y+\frac{\hbar \omega _{c}}{\Gamma }\left( \frac{%
\omega }{\omega _{c}}-n\right) \right] ^{2}\right\} dy .  \label{eq31}
\end{equation}%
It is obvious that $J_{2,1}\left( \omega /\omega _{c}\right) $ has prominent
maxima near the conditions $\omega /\omega _{c}=1,2,...$, if the 2D electron
system is pure enough and $\hbar \omega _{c}/\Gamma >1$. The results of
numerical evaluations of the function $J_{2,1}\left( \omega /\omega
_{c}\right) $ and its derivative $J_{2,1}^{\prime }\left( \omega /\omega
_{c}\right) $ are shown in Fig.~\ref{f5} by the solid and dashed (dashed-dotted)
lines respectively. The calculations were performed for $N_{2}=0.1N_{e}$ and two values of the
magnetic field [$B=0.5\,\mathrm{T}$ (red lines) and $0.1\,\mathrm{T}$ (blue
lines)]. The heights of the maxima increase with lowering $B$ due to the
factor $\left( 1-\eta _{2}\right) /\eta _{2}$ because the filling factor $%
\eta _{2}\left( 0.5\,\mathrm{T}\right) \simeq 0.165$ while $\eta _{2}\left( 0.1%
\,\mathrm{T}\right) \simeq 0.827$. The change of $B$ affects notably also the
positions of minima, maxima and the zero-crossing (sign-changing) point of
the derivative $J_{2,1}^{\prime }\left( \omega /\omega _{c}\right) $.

Using the same approximations as those used for obtaining Eq.~(\ref{eq31}),
the abnormal contribution to the effective collision frequency can be
represented as%
\begin{equation}
\nu _{\mathrm{A}}=\nu _{0}\frac{2p_{2,1}}{\pi \eta }\left( \frac{\hbar
\omega _{c}}{\Gamma }\right) ^{2}\left( 1-\mathrm{e}^{-\mu
_{2,1}/T_{e}}\right) \Phi _{2,1}\left( B\right) \text{ },  \label{eq32}
\end{equation}%
where we defined \
\begin{equation}
\Phi _{2,1}\left( B\right) =\frac{\Gamma }{\hbar \omega _{c}}J_{2,1}^{\prime
}\left( \omega _{2,1}/\omega _{c}\right)   \label{eq33}
\end{equation}%
because the derivative $J_{2,1}^{\prime }\left( \omega _{2,1}/\omega
_{c}\right) $ contains the additional factor $\hbar \omega _{c}/\Gamma $
according to Eq.~(\ref{eq31}). The dimensionless parameter $p_{l,l^{\prime }}$
is determined by the following matrix elements
\begin{equation}
p_{l,l^{\prime }}=\frac{B_{1,1}}{B_{l,l^{\prime }}},\text{ \ \ }%
B_{l,l^{\prime }}^{-1}=L_{z}^{-1}\sum_{\kappa }\left\vert \left( e^{-i\kappa
z}\right) _{l^{\prime },l}\right\vert ^{2}\text{ }.  \label{eq34}
\end{equation}%
The accurate calculation of $p_{2,1}$ requires
the knowledge of the details of a particular 2D electron system such as the
wavefunctions of subband states which are not considering in this
work. For electrons on liquid helium~\cite{KonMonKon-2007}, $p_{2,1}$ is a
factor of two smaller than $p_{1,1}=1$. Therefore, in following numerical calculations
we shall use a rough estimation: $2p_{2,1}\simeq 1$.

Under the conditions used for obtaining Eq.~(\ref{eq32}), the contribution from
electron scattering within the ground subband ($l=1$) can be found as
\begin{equation}
\nu _{\mathrm{intra}}^{(1)}\simeq \nu _{0}\frac{p_{1,1}}{\pi \eta }\left( \frac{%
\hbar \omega _{c}}{\Gamma }\right) ^{2}\Phi _{1,1}\left( B\right) \text{ ,}
\label{eq35}
\end{equation}%
where%
\begin{equation}
\Phi _{1,1}\left( B\right) =\sum_{n=0}^{\infty }\left( 2n+1\right) \exp
\left[ -\frac{4\left( \mu _{1}-\varepsilon _{n}\right) ^{2}}{\Gamma ^{2}}%
\right] \text{ }.  \label{eq36}
\end{equation}%
At the same time, the  contribution from electron scattering within the first
excited subband $\nu _{\mathrm{intra}}^{(2)}$ has a very weak dependence on $B$ because
the distribution function $f_{2}\left(\varepsilon \right)$ given in Eq.~(\ref{eq29})
varies strongly near $\varepsilon _{0}$. Thus, $\nu _{\mathrm{intra}}^{(2)}$ can
be considered as a small background value when the ratio $N_{2}/N_{e}\ll 1$.
The background value decreases also with narrowing of the density of states.
In the following, we shall neglect $\nu _{\mathrm{intra}}^{(2)}$ and assume that
$\nu _{\mathrm{intra}}\simeq \nu _{\mathrm{intra}}^{(1)}$.

Comparing $\nu _{\mathrm{A}}$ of Eq.~(\ref{eq32}) with $\nu _{\mathrm{intra}}^{(1)}$
given in Eq.~(\ref{eq35}) indicates that the abnormal contribution
contains the additional factor $\left( 1-\mathrm{e}^{-\mu
_{2,1}/T_{e}}\right) $ which is zero under equilibrium conditions ($\mu
_{2}=\mu _{1}=\mu _{\mathrm{F}}$). If we have a nonequilibrium population of
the second subband, then, according to Eq.~(\ref{eq27}) and Fig.~\ref{f2}, $%
\delta \mu _{2}$ becomes substantially larger than $T_{e}$ already at a small
filling factor $\eta _{2}$. For example, Fig.~\ref{f2} shows that $\mu
_{2}>\Delta _{2,1}$ if $\eta _{2}>0.1$. Assuming this reasonable condition,
we can neglect the exponentially small term in the factor $\left( 1-\mathrm{e%
}^{-\mu _{2,1}/T_{e}}\right) $ and set this factor to unity even if $\mu _{1}
$ is fixed (according to Fig.~\ref{f3}, $\mu _{1}$ decreases with lowering $%
N_{1}$ which also reduces $\mathrm{e}^{-\mu _{2,1}/T_{e}}$).

\begin{figure}[tbp]
\begin{center}
\includegraphics[width=10.0cm]{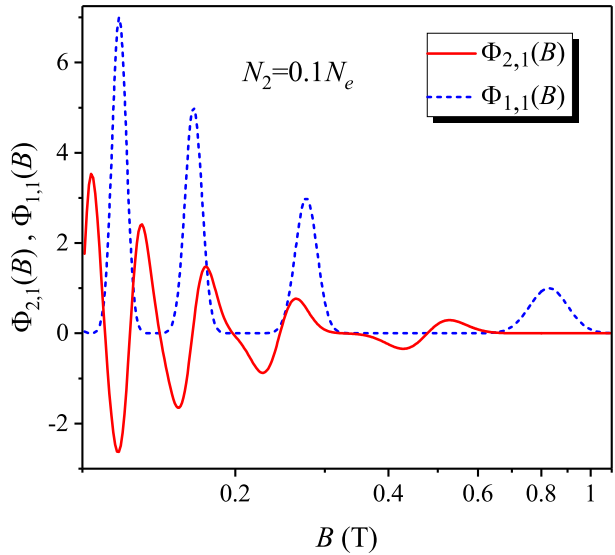}
\end{center}
\caption{Graphical illustration of the functions $\Phi _{2,1}(B)$ and $\Phi _{1,1}(B)$ which
determine $\nu _{\mathrm{A}}$ and $\nu _{\mathrm{intra}}$ respectively.
} \label{f6}
\end{figure}

Another important distinction between $\nu _{\mathrm{A}}$ and $\nu _{\mathrm{%
intra}}$ is caused by different behaviors of the dimensionless functions $%
\Phi _{2,1}\left( B\right) $ and $\Phi _{1,1}\left( B\right) $ illustrated
in Fig.~\ref{f6}. The both functions oscillate with varying $1/B$, but the
periods of these oscillations are different. Assuming $\mu _{1}$ is fixed to
$\mu _{\mathrm{F}}$, the maxima of the positive function $\Phi _{1,1}\left(
B\right) $ entering $\nu _{\mathrm{intra}}$ occur at $\hbar \omega _{c}=\mu
_{\mathrm{F}}/\left( n+1/2\right) $ due to the Shubnikov--de Haas effect. In
contrast to $\Phi _{1,1}\left( B\right) $, the function $\Phi _{2,1}\left(
B\right) $, which determines $\nu _{\mathrm{A}}$, is a sign-changing function
having maxima and minima, according to the definition of Eq.~(\ref{eq33}) and
Fig.~\ref{f5}; its zero-crossing points occur at magnetic fields which are
close to the condition $\Delta _{2,1}/\hbar \omega _{c}=m$ (here $m=1,2,...$%
).

It is instructive to analyze $\nu _{\mathrm{N}}$ using the same
approximations and conditions.
Direct transformation of Eq.~(\ref{eq25}) yields
\begin{equation}
\nu _{\mathrm{N}}=\nu _{0}\frac{2p_{2,1}}{\pi \eta }\mathrm{e}^{-\mu _{2,1}/T_{e}}
\frac{\hbar ^{2}\omega _{c}^{2}}{T_{e}\Gamma }J_{2,1}\left(
\omega _{2,1}/\omega _{c}\right)  .  \label{eq37}
\end{equation}%
As compared to the contribution from intrasubband scattering of Eq.~(\ref{eq35}),
here we have $T_{e}$ in the denominator because for intersubband scattering one
cannot use the relationship
 $f\left( \varepsilon \right) \left[ 1-f\left( \varepsilon \right) \right]
\rightarrow T_{e}\delta \left( \varepsilon -\varepsilon _{\mathrm{F}}\right)
$. The shape of oscillations caused by $\nu _{\mathrm{N}}$ is determined by
the function $J_{2,1}\left( \omega _{2,1}/\omega _{c}\right)$ shown above in
Fig.~\ref{f5} by solid lines. This shape is in a qualitative accordance with
results obtained for magnetointersubband oscillations under equilibrium conditions~\cite{RaiSha-1994}.
For nonequilibrium regime described here, Eq.~(\ref{eq37}) contains also
the exponential factor $\exp \left( -\mu _{2,1}/T_{e} \right) $
which becomes very small even for relatively weak excitations $N_{2}=0.1N_{e}$.
It should be noted also that under conditions used here, the amplitude of
$\Phi _{2,1}$ is about 5 times larger than the respective amplitude of $J_{2,1}$.
Therefore, $\nu _{\mathrm{N}}$ can be neglected as compared to $\nu _{\mathrm{A}}$ and
$\nu _{\mathrm{intra}}$.

\begin{figure}[t]
\begin{center}
\includegraphics[width=10.0cm]{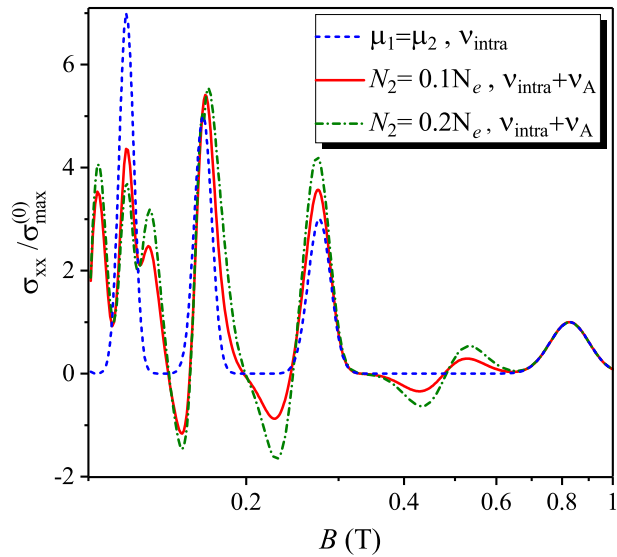}
\end{center}
\caption{Magnetoconductivity normalized to $\sigma _{\max }^{\left( 0\right) }=e^{2}/4\pi \hbar $
versus the magnetic field for different levels of the displacement from equilibrium: $N_{2}\simeq 0$ (blue dashed line),
$N_{2}=0.1N_{e}$ (red solid line), and $N_{2}=0.2N_{e}$ (olive dash-dotted line).
} \label{f7}
\end{figure}

Typical dependencies of $\sigma _{xx}\left(B\right)$ are
shown in Fig.~\ref{f7}. In the equilibrium case ($\mu _{1}=\mu _{2}$), $\nu _{%
\mathrm{eff}}=\nu _{\mathrm{intra}}$ and $\sigma _{xx}\left( B\right) $ has
maxima when $\hbar \omega _{c}=\mu _{\mathrm{F}}/\left( n+1/2\right) $
according to the SCBA theory~\cite{AndUem-1974} (blue dashed line). In this
figure, the electron conductivity $\sigma _{xx}$ is normalized by the first (%
$n=0$) peak value $\sigma _{\max }^{\left( 0\right) }=e^{2}/4\pi \hbar $
found for the Gaussian level density ($B\simeq 0.827\,\mathrm{T}$).
Already a small nonequilibrium electron population of
the excited subband ($N_{2}/N_{e}=0.1$) induces
important changes into $\sigma _{xx}\left( B\right) $ shown in Fig.~\ref{f7}
by the red line. Besides additional maxima and a substantial reduction of the
SCBA peak at $n=3$, there are sign-changing variations of $\sigma
_{xx}\left( B\right) $ near $B\simeq 0.48\,\mathrm{T}$, $0.24\,\mathrm{T}$ and $%
0.156\,\mathrm{T}$, and quite deep minima with regions where the linear response
conductivity $\sigma _{xx}$ becomes negative. An increase in the electron
population of the excited subband ($N_{2}/N_{e}=0.2$) amplifies these
unusual phenomena as indicated in Fig.~\ref{f7} by the olive dash-dotted line.
It should be noted that for such a population, $\eta _{2}\left( B\right) $
becomes larger than unity in the region of low $B$, and, therefore, the
approximation of Eq.~(\ref{eq27}) defining $\mu _{2}$ fails. In this case, we
had used the extension of the quasi-Fermi energy given in Eq.~(\ref{eq28}).
Numerical calculations indicate also that reducing temperature from $1%
\,\mathrm{K}$ to $0.5\,\mathrm{K}$ amplifies additionally the effect of the
sign-changing contribution $\nu _{\mathrm{A}}$.

Thus, the theoretical analysis given above indicates that the Pauli exclusion principle
does not ruin the intersubband mechanism of MIRO, if the electron distribution
in the ground and excited subbands can be described by the quasi-Fermi level approximation.
Moreover, a sharp increase of the imref of the excited subband as a function of the filling
factor shown in Fig.~\ref{f2} reduces strongly the compensational contribution from reverse
intersubband scattering [the exponential term in parenthesis of Eq.~(\ref{eq32});
under conditions of Fig.~\ref{f7} this term does not exceed 0.04]. This means that
magnetoconductivity oscillations and ZRS induced by the resonant MW field, whose polarization direction is
perpendicular to the electron layer, can be realized in sufficiently clean semiconductor devices.
The regions with negative linear response conductivity attract a special interest, because they
allow performing complementary studies of ZRS in heterostructures caused by a definite
mechanism. These studies potentially can help also with the identification of the origin of MIRO
and ZRS in the conventional setup.

\section{Conclusion}

We have presented a theory of quantum magnetotransport in a degenerate
multisubband electron system under conditions that electron distributions
over 2D subbands cannot be described by a single chemical
potential. Using the concept of quasi-Fermi levels and the
self-consistent Born approximation, we expressed magnetoconductivity equations
in terms of the extended dynamic structure factor and its derivative
with regard to frequency. We have shown that a displacement from the equilibrium
electron distribution over excited subbands, which cannot be reduced to trivial
heating, leads to appearance of abnormal sign-changing contribution to the
momentum collision rate and magnetoconductivity. Calculations performed for a
simplified potential of scatterers indicate that even a small fraction of electrons
(about 10\%) transferred to the first excited subband can drastically change the shape of
magnetointersubband oscillations an lead to negative linear response conductivity.
The theory can be applied to electrons on helium films with a special arrangements of
substrates~\cite{PeePla-1983}, and to multisubband 2D electron systems of semiconductor devices.

\end{document}